\documentclass[12pt,a4paper]{article}

\usepackage[includeheadfoot,
            marginratio={1:1,2:3}, 
            width=450pt, 
            height=688pt,]{geometry}

\usepackage{verbatim}
\usepackage{amsmath}
\usepackage{amsfonts}
\usepackage{amssymb}
\usepackage{mathtools}
\usepackage{physics}
\usepackage{siunitx}
\usepackage{graphicx}
\usepackage{tikz} 
\usetikzlibrary{decorations.pathreplacing,calc,calligraphy}
\usepackage{booktabs}
\usepackage{adjustbox}
\usepackage[utf8]{inputenc}
\usepackage{multirow}
\usepackage[splitrule]{footmisc}

\usepackage{empheq}
\usepackage{paralist}
\usepackage[hidelinks]{hyperref}
\usepackage{xcolor}
\usepackage{tensor}

\usepackage{nicefrac}

\newcommand{\nc}{\newcommand}
\newcommand\blfootnote[1]{%
  \begingroup
  \renewcommand\thefootnote{}\footnote{#1}%
  \addtocounter{footnote}{-1}%
  \endgroup
}
\nc{\lb}{\llbracket}
\nc{\rb}{\rrbracket}
\nc{\gl}{\llbracket}
\nc{\gr}{\rrbracket}
\nc{\del}{\partial}
\nc{\tri}{\hspace{-3.5pt}\vartriangle\hspace{-3.5pt}}
\nc{\blacktri}{\blacktriangle}

\nc{\eq}[1]{\begin{equation}
                     \begin{split} #1 \end{split}
                     \end{equation}}
\nc{\ul}{\underline}
\nc{\ov}{\overline}

\nc{\fa}{\hat}
\nc{\fb}{\MakeUppercase}
\nc{\fc}{\tilde}
\nc{\Lie}{{\cal L}} 
\nc{\lambdabar}{{\mkern0.75mu\mathchar '26\mkern -9.75mu\lambda}}

\allowdisplaybreaks[2]
\numberwithin{equation}{section}
\newcommand{\Lsp}{\ensuremath{\Lambda_\text{sp}}}

\DeclareUnicodeCharacter{2212}{-}
\begin{document}

\vspace*{-1.5cm}
\begin{flushright}
  {\small
  MPP-2025-201\\
  }
\end{flushright}

\vspace{0.5cm}
\begin{center}
  {\Large \bf
Higher Curvature Inflation and the Species Scale
} 
\vspace{0.2cm}

\end{center}

\vspace{0.15cm}
\begin{center}
Joaquin Masias\\[0.2cm]
\blfootnote{jmasias@mpp.mpg.de}
\end{center}

\vspace{0.0cm}
\begin{center} 
{\footnotesize
\vspace{0.25cm} 
\emph{Max-Planck-Institut für Physik (Werner-Heisenberg-Institut),\\ Boltzmannstr. 8, 85748 Garching, Germany}\\[0.1cm]
\vspace{0.25cm} 
}
\end{center} 

\vspace{0.3cm}

\begin{abstract}
\noindent 
We study the scalar potentials that arise from higher curvature corrections in general $f(R)$ theories of gravity and their connection to a dynamical species scale. Starting from general considerations in arbitrary dimensions, we show that at large field values, the scalar potential generated by an infinite series of curvature terms and the field-dependent species scale arising from circle compactification both decay exponentially, in complementary ways. We identify a direction in field space along which these two effects cancel, giving rise to exponentially flat, plateau-like potentials, and show that a plateau of cosmological duration requires this direction to be fixed by an additional sector. Additionally we discuss possible implications the mechanism proposed could have for M-theory.
\end{abstract}

\clearpage

\section{Introduction}

Cosmological inflation provides a compelling mechanism for addressing several phenomenological issues of standard cosmology, such as the horizon and flatness problems, while simultaneously generating nearly scale-invariant primordial fluctuations in agreement with CMB data \cite{Planck:2018vyg, BICEP:2021xfz}. Among the broad landscape of inflationary models, the Starobinsky model \cite{Starobinsky:1980te} remains one of the most robust and elegant scenarios. In its simplest form, the addition of an $R^2$ term to the Einstein-Hilbert action gives rise to a scalar degree of freedom whose potential naturally exhibits a plateau at large field, as well as a Minkowski minimum around which the inflaton can enter a reheating phase. The resulting predictions for the spectral tilt $n_s$ are in remarkable agreement with data, and the predicted values for the tensor-to-scalar ratio $r$ are within reach of the next generation of B-mode polarization experiments \cite{Abazajian:2019eic,LiteBIRD:2024wix}. The Starobinsky model naturally connects to the broader class of 
$f(R)$ and more general modified theories of gravity. These are often studied as extensions of general relativity from an effective field theory (EFT) perspective (see e.g. \cite{Ferrara:2013kca, Huang:2013hsb} for applications to cosmology, and \cite{Sotiriou:2008rp} for a review).

 It has long been understood that a renormalizable theory of gravity requires extending the Einstein-Hilbert action by higher curvature operators \cite{Utiyama:1962sn, Stelle:1977ry}. From the Wilsonian EFT perspective, the $R^2$ term should not be isolated, but rather part of an infinite tower of higher-derivative terms \cite{Donoghue:1994dn}. An immediate question is then whether the inflationary plateau can remain once such corrections are included.

This is closely tied to our recent understanding of quantum gravity in the context of the Swampland program, which attempts to distinguish EFTs consistent with quantum gravity from those which are not \cite{Vafa:2005ui}. In particular, the Swampland Distance Conjecture (SDC) \cite{Ooguri:2006in} (and its refinement \cite{Lee:2019wij}) suggests that large scalar field excursions are necessarily accompanied by towers of light states, which invalidate the naive low-energy effective description. The effects of coupling such towers to the inflaton in models of large-field inflation were studied in \cite{Lust:2025auk}.
A key role in this is played by the \emph{species scale} $\Lsp$ \cite{Dvali:2007hz, Dvali:2007wp, Dvali:2010vm}, which sets the effective cutoff of the theory in the presence of a large number of light species, and, as has been recently explored in the literature \cite{vandeHeisteeg:2022btw, vandeHeisteeg:2023dlw, Castellano:2023aum}, can be identified as the scale suppressing higher curvature corrections to any EFT coupled to gravity. The species scale can generally be a field-dependent quantity \cite{vandeHeisteeg:2022btw}, for instance in Kaluza-Klein compactifications, where it depends exponentially on the volume modulus of the compact space.

The role of the Swampland program in Starobinsky inflation was also studied in \cite{Lust:2023zql} (see also \cite{brinkmann2023starobinsky}), where it was argued that, in UV-complete theories of gravity, the $R^2$ term should necessarily be weighted by the quantum gravitational cutoff. This suggests that inflation would occur at energy scales where the weakly coupled EFT cannot be trusted. Moreover, as large field displacements occur during inflation, the scalar potential itself is expected to vary significantly, thereby destabilizing the plateau even in the absence of explicit higher curvature corrections.

In this work, we find that at large field values, the species scale coming from compactification on a circle and the scalar potential coming from an infinite curvature series may decay exponentially in complementary ways. We take this as motivation to study how the interplay of higher curvature corrections and the species scale can contribute to generating plateau-like scalar potentials and whether they can be made consistent with Swampland
constraints.

The structure of this paper is as follows. In Section~\ref{sec:ii} we revisit the construction of scalar potentials from general $f(R)$ theories in arbitrary dimension. We recall how plateau-like potentials can only emerge for specific choices of asymptotics, and generically one has exponentially decaying potentials. In Section 3 we compactify on a circle and study the two field dynamics of the radion and the scalaron. There is a direction in field space along which the exponential growth of the cutoff cancels the decay of the potential. We discuss how this direction can be fixed by a combination of localized and bulk terms, and whether this can be a stable configuration. We then analyze the resulting plateau potentials. In Section~\ref{sec:iv} we discuss the possible implications for inflationary cosmology and for M-theory. Finally, in Section~\ref{sec:v} we conclude.

\section{Scalar Potentials from \texorpdfstring{$f(R)$}{f(R)} Gravity}
\label{sec:ii}

We start by reviewing how modified gravity theories of the form $f(R)$ in $d$ dimensions are recast as Einstein-Hilbert gravity coupled to an additional scalar degree of freedom\footnote{In Appendix \ref{sec:appi} we also show that during a quasi-de Sitter phase one can neglect other types of curvature invariants.}.

Throughout we restrict ourselves to actions built out of the Ricci scalar alone. This is an assumption on the class of theories we study, and not a statement about the generic form of a gravitational effective action, which contains every curvature invariant, nor about the corrections that appear in UV complete theories of gravity, such as string theory, which are generically not of the Ricci type. In Appendix \ref{sec:appi} we argue that on a nearly maximally symmetric background the remaining invariants reduce to powers of $R$, so that the restriction is consistent on the backgrounds we consider. We also take the curvature series to have a finite radius of convergence, which is again an assumption. Series whose coefficients grow factorially are not generally considered here, and we only briefly discuss them in Section \ref{sec:ivb} and Appendix \ref{sec:appii}.

Consider the action for $f(R)$ gravity:
\begin{equation}
    S = \frac{M_{\text{Pl}}^{d-2}}{2} \int d^d x \sqrt{-g} f(R)\,,
\end{equation}
where $R$ denotes the scalar curvature in the Jordan-frame, as opposed to Einstein-frame. We can write the action in the form:
\begin{equation}
    S = \int d^d x \sqrt{-g} \left[ \frac{M_{\text{Pl}}^{d-2}}{2} F R - U \right],
\end{equation}
where the functions $F$ and $U$ are defined as
\begin{equation}
    F = \frac{\partial f}{\partial R}, \quad
    U = \frac{M_{\text{Pl}}^{d-2}}{2} (F R - f).
\end{equation}

To proceed to the Einstein frame, we perform a Weyl rescaling of the metric:
\begin{equation}
    g_{\mu\nu} \rightarrow \tilde{g}_{\mu\nu} = \Omega^2 g_{\mu\nu} = e^{2\varphi} g_{\mu\nu},
\end{equation}
under which the determinant and Ricci scalar transform as
\begin{align}
    \sqrt{-g} &= \Omega^{-d} \sqrt{-\tilde{g}}, \\
    R &= \Omega^2 \left( \tilde{R} + 2(d-1)\tilde{\Box} \varphi - (d-1)(d-2) \tilde{g}^{\mu\nu} \partial_\mu \varphi \partial_\nu \varphi \right).
\end{align}
Choosing the conformal factor as
\begin{equation}
    F = \Omega^{d-2},
\end{equation}
one obtains the Einstein-frame action
\begin{equation}
    S = \int d^d x \sqrt{-\tilde{g}} \left[ \frac{M_{\text{Pl}}^{d-2}}{2} \tilde{R} - \frac{1}{2} \tilde{g}^{\mu\nu} \partial_\mu \sigma \partial_\nu \sigma - V(\sigma) \right],
\end{equation}
such that $\tilde{R}$ is the Einstein-frame scalar curvature. Here we have defined
\begin{equation}
    \sigma = \sqrt{(d-1)(d-2)}\, M_{\text{Pl}}^{\frac{d-2}{2}}\, \varphi\,, 
\end{equation}
such that
\begin{equation}
    V(\sigma) = \frac{U}{F^{d/(d-2)}}, \quad F = e^{\sqrt{\frac{d-2}{d-1}} \frac{\sigma}{M_{\text{Pl}}^{\frac{d-2}{2}}}}.
\end{equation}
While $F$ can always be written explicitly in terms of $\sigma$, obtaining an analytic form of $V(\sigma)$ is generally only feasible for specific cases of $f(R)$ \cite{Broy:2015zba}. As we will attempt to connect the two, we briefly discuss asymptotically flat and exponentially decaying potentials arising from this sort of theory.

\subsection{Asymptotically Flat Potentials from \texorpdfstring{$f(R)$}{f(R)} Gravity}

To identify the functional forms of $f(R)$ that yield asymptotically flat scalar potentials, we analyze the large-field behavior of $V(\sigma)$ in the limit $\sigma \to \infty$. In this regime, we demand that
\begin{equation}
    \lim_{\sigma \to \infty} V(\sigma) = c_1, \quad \lim_{\sigma \to \infty} (F R - f) = c_2 F^{\frac{d}{d-2}},
\end{equation}
\begin{equation}
    \lim_{\sigma \to \infty} \frac{dV}{d\sigma} = 0, \quad \lim_{\sigma \to \infty} \left[ \frac{R}{F^{2/(d-2)}} \left( \frac{2}{d} - \frac{f}{R F} \right) \right] = 0,
\end{equation}
for constants $c_1, c_2 > 0$. Both right hand sides follow from the definition of $V$ and the fact that
\begin{equation}
    \frac{d}{dF}\left(FR-f\right)=R\,.
\end{equation}
The first is then immediate, since $V=\tfrac{1}{2}M_{\text{Pl}}^{d-2}(FR-f)F^{-d/(d-2)}$ tends to a constant only if $FR-f$ grows as $F^{d/(d-2)}$. For the second, $F$ is exponential in $\sigma$, so that $dV/d\sigma\propto F\,dV/dF$ and
\begin{equation}
    F\frac{dV}{dF}\,=\,-\frac{d}{d-2}\,\frac{M_{\text{Pl}}^{d-2}}{2}\,\frac{R}{F^{2/(d-2)}}\left(\frac{2}{d}-\frac{f}{RF}\right)\,.
\end{equation}
Imposing both conditions then requires $f/(RF)\to2/d$ at large $R$, and integrating $f'/f=d/(2R)$ gives \cite{Broy:2014xwa,Broy:2015zba}
\begin{equation}
    f(R) \propto R^{d/2}.
\end{equation}

A minimal realization of this is
\begin{equation}
    f(R) = R + \frac{R^{d/2}}{M^{d-2}}\,,
\end{equation}
with $M$ some energy scale. More generally, any function of the form
\begin{equation}
    f(R) = R + \frac{R^{d/2}}{M^{d-2}} g(R)\,,
\end{equation}
where $g(R) \to \text{const} > 0$ as $R \to \infty$, leads to an asymptotically flat potential.

As a concrete example, consider a scalar potential with an exponential plateau
\begin{equation}
    V(\sigma) \simeq V_0 \left(1 - e^{-\gamma \sqrt{\frac{d-2}{d-1}} \sigma} \right).
\end{equation}
This corresponds, in the large-$\sigma$ limit, to an $f(R)$ theory of the form:
\begin{equation}
    f(R) \simeq R^{d/2} + R^{1 - \frac{d-2}{2}(\gamma - 1)}\,,
\end{equation}
where, only for $\gamma = 1$, one has a theory with an Einstein-Hilbert term in the Jordan frame. 
\subsection{Exponential Potentials from \texorpdfstring{$f(R)$}{f(R)} Gravity}
When the leading term in the curvature expansion is of order $n$
\begin{equation}
    f(R)=R+M_0^2 (R/M_0^2)^n\,, 
\end{equation}
we will have an exponential potential of the form

\begin{equation}
V(\sigma) = \frac{1}{2} M_0^2 M_{\text{Pl}}^{d-2} (n-1) \left( \frac{1}{n} \right)^{\frac{n}{n-1}} e^{- \sqrt{\frac{d^2}{(d-2)(d-1)}} \sigma} \left( e^{\sqrt{\frac{d-2}{d-1}} \sigma} - 1 \right)^{\frac{n}{n-1}}.
\end{equation}
At large $\sigma$ we have
\begin{equation}
V(\sigma) \simeq \frac{1}{2} M_0^2 M_{\text{Pl}}^{d-2} (n-1) \left( \frac{1}{n} \right)^{\frac{n}{n-1}} e^{-\frac{1}{\sqrt{(d-2)(d-1)}}\frac{2 n-d}{n-1} \sigma} \left(1- \frac{n}{n-1}e^{-\sqrt{\frac{d-2}{d-1}} \sigma} \right).
\end{equation}
and the potential decays exponentially at a rate
\begin{equation}
\label{eq:gammaexp}
   2\, \gamma=\frac{1}{\sqrt{(d-2)(d-1)}}\frac{2 n-d}{(n-1)}.
\end{equation}
For integer $n<d/2$ we simply have that the potential is exponentially bounded from both sides. For $n>d/2$, $\gamma$ satisfies
\begin{equation}
\qquad\quad\qquad0\leq\gamma\leq \dfrac{1}{\sqrt{(d-2)(d-1)}}\,,
\end{equation}
with the upper bound saturated when $n\to\infty$. We can argue that this maximum decay rate will be generic for infinite series. To this end, consider a potential characterized by two exponential scales:
\begin{equation}
\label{eq:poteqs}
    V(\sigma) \simeq \frac{M_0^2 M_{\text{Pl}}^{d-2}}{2} e^{-2\gamma \sigma} \left( 1 - a_0 e^{-\lambda \sigma} \right)\,,
\end{equation}
coming from a curvature series 
\begin{equation}
    f(R)=\sum_{n=1} c_n M_0^2 \left(\dfrac{R}{M_0^2}\right)^n.
\end{equation}
The exponential parameter $\gamma$ is related to the asymptotic form of $f(R)$ via
\begin{equation}
\label{eq:gammaMaster}
    \gamma = -\frac{1}{2} \left( \frac{1}{V} \frac{dV}{d\sigma} \right)_{\sigma \to \infty} = \frac{1}{\sqrt{(d-2)(d-1)}} \frac{1 - \frac{d}{2} \frac{f}{R F}}{1 - \frac{f}{R F}} \Bigg|_{R \to M_0^2}\,,
\end{equation}
where we have assumed that the infinite distance limit of $\sigma$ corresponds to $R\to M_0^2$. This is equivalent to demanding that $F$ diverges at $M_0^2$
\begin{equation}
\sum_{n=1} c_n\,n=\infty.
\end{equation}
Given this, and assuming that the series is such that its partial sums are bounded from below\footnote{This condition is, for example, weaker than requiring the series to contain only a finite number of negative coefficients, and it is further supported by positivity constraints on irrelevant operators \cite{Adams:2006sv}.}, 
\begin{equation}
    \sum_{n=1}^N c_n > -\infty \,, \quad \forall N,
\end{equation}
it follows that

\begin{equation}
    \lim_{R\to M^2_0} \dfrac{f}{R F}\to 0\,,
\end{equation}
and 
\begin{equation}
\label{eq:gammares}
    \gamma =
        \dfrac{1}{\sqrt{(d-2)(d-1)}}.
\end{equation}
As such, under some assumptions we obtain that an infinite series of curvature corrections leads to an exponentially decaying scalar potential with the same decay rate as one would obtain from compactification on a circle \cite{vandeHeisteeg:2023ubh}, independently of the number of dimensions. 
For rational coefficients, if we assume some asymptotic spectrum $c_n = n^q$, one can explicitly check:
\begin{equation}
    \gamma =
        \dfrac{1}{\sqrt{(d-2)(d-1)}}, \quad{} \text{for } q \geq -2,
\end{equation}
since for $q<-2$, $F$ is convergent at $R=M_0^2$.

To determine the second exponent $\lambda$, we define the dimensionless rescaled potential
\begin{equation}
    \tilde{V} = \frac{2\,e^{2\gamma \sigma} V(\sigma)}{M_0^2 M_{\text{Pl}}^{d-2}}\,,
\end{equation}
and compute
\begin{equation}
\label{eq:lambdaMaster}
    \lambda = \left. \frac{d\tilde{V}/d\sigma}{1 - \tilde{V}} \right|_{\sigma \to \infty} = \left. \frac{\sqrt{\frac{d-2}{d-1}}}{1 + (M_0^2 - R) \frac{F}{f}} \right|_{R \to M_0^2}.
\end{equation}

For a monomial spectrum $c_n = n^q$ this is:
\begin{equation}
\label{eq:lambdares}
    \lambda =
    \begin{cases}
        \sqrt{\frac{d-2}{d-1}} \cdot \frac{1}{2 + q} & \text{for } q \geq -1\,, \\
        \sqrt{\frac{d-2}{d-1}} & \text{for } -2 < q \leq -1\,.
    \end{cases}
\end{equation}
Then $\lambda$ is upper-bounded by $\sqrt{\frac{d-2}{d-1}}$, and will saturate this bound whenever the coefficients satisfy $1/n^2\leq c_n\leq1/n$. Apart from these, we also briefly discuss other types of curvature series in Appendix \ref{sec:appii}.\footnote{We expect the coefficients to be constrained by the analyticity and unitarity of graviton scattering amplitudes as in \cite{Caron-Huot:2024lbf}, $c_n^{\frac{1}{2n-2}}M\lesssim \Lsp$, so the coefficients cannot be arbitrarily large. In the limit $M \simeq \Lambda_{\mathrm{sp}}$, we expect them to be strictly non-increasing.}

 In an EFT approach, and also from the typical expectation in string theory, one expects that the scale controlling the higher-order terms in the series corresponds to the UV cutoff of the series. This cutoff has been identified in various contexts as the species scale $\Lambda_{\text{sp}}=M_{\text{Pl}}/N_{\text{sp}}^{\frac{1}{d-2}}$, where $N_{\text{sp}}$ is the number of light degrees of freedom in the theory. This is generally a moduli-dependent quantity, but has been conjectured to also hold on a general scalar field space\footnote{Although there is no unifying picture for the SDC in a general field space, this has been studied from several different perspectives (see e.g. \cite{Klaewer:2016kiy, Baume:2016psm, Basile:2023rvm, Demulder:2023vlo,  Debusschere:2024rmi, Demulder:2024glx, Mohseni:2024njl}).} \cite{Palti:2019pca}. It is then somewhat remarkable that potentials with exponential decay rates given by \eqref{eq:gammares} are exactly what one would obtain from compactifying a theory one dimension higher with a cosmological constant $\Lambda = M_0^2/2$ on a circle. Also in that same setup, the species scale after compactification will satisfy $\Lambda_{\text{sp}} \simeq \sqrt{2V(\sigma)}$.
Motivated by this, we explore the relation further, to try and obtain asymptotically flat potentials.

\section{\texorpdfstring{$f(R)$}{f(R)} gravity on a circle}
\label{sec:iii}
In this section we study how simple Kaluza-Klein (KK) reductions on a circle, $x^d\sim x^d+2\pi r$, affect the $f(R)$ action in the lower-dimensional theory, and how the resulting radion dependence can naturally induce a field-dependent cutoff (or species) scale. The lower-dimensional effective theory takes the form of an $f(R)$ model in which the Jordan-frame action is coupled non-minimally to a scalar field. After reduction on the circle we have two scalar degrees of freedom, the radion and the scalaron. This is expected of a higher curvature theory coupled non-minimally to a scalar, which has two tensor and two scalar modes, provided the background is not flat and the map between the Jordan and Einstein frames is non-singular \cite{Hell:2023mph,Hell:2025lbl}. Consider the $D=d+1$-dimensional action
\begin{equation}
\label{eq:SD}
S_D = \frac{M_{\text{Pl},D}^{D-2}}{2} \int d^D x \sqrt{-G} f(R_D)=\frac{M_{\text{Pl},D}^{D-2}}{2} \int d^D x \sqrt{-G} \left[ R_D + M_0^2 \sum_{n=2} c_n \frac{R_D^n}{M_0^{2n}} \right],
\end{equation}
where $M_0$ is the mass scale controlling higher curvature corrections, $M_{\text{Pl},D}$ is the higher-dimensional Planck mass, and $c_n$ are dimensionless coefficients.

For the dimensional reduction we use the standard metric ansatz
\begin{equation}
\label{eq:ansatz}
ds^2 = G_{MN} dx^M dx^N = e^{2\alpha\rho} g_{\mu\nu} dx^\mu dx^\nu + e^{2\beta\rho} (dx^d)^2,
\end{equation}
with the radion, $\rho$, independent of the internal coordinate $x^d$.
We will consider (nearly) maximally symmetric solutions, such that the universe is in a quasi-de Sitter phase. The slow-roll parameter is defined as
\begin{equation}
    \varepsilon\equiv -\dfrac{\dot H}{H^2}\,,
\end{equation}
where $H$ is the Hubble parameter. Then, being in a slow-roll phase requires
\begin{equation}
    \varepsilon\ll 1 \,,
\end{equation}
so that we in fact have quasi-de Sitter.
This allows us to neglect higher-derivative kinetic terms in the scalar field \footnote{During slow-roll we have $\dot\rho^2\simeq -(d-2)\dot H\ll H^2$, so higher derivatives of $\rho$ can be neglected.}. A canonical scalar kinetic term in $d$ dimensions is obtained by choosing
\begin{equation}
\label{eq:alphabeta}
\alpha^2 = \frac{1}{(d-1)(d-2)}, \qquad \beta = - (d - 2) \alpha\,,
\end{equation}
which fixes the radion normalization.

After performing the reduction on the circle one obtains 
\begin{equation}
\label{eq:RD}
R_D = e^{-2\alpha\rho}\Big[\,R - 2\alpha\,\partial^2 \rho - (\partial\rho)^2\,\Big]\,,
\end{equation}
where $R$ and $\partial$ are defined with respect to $g_{\mu\nu}$. Together with $\sqrt{-G}=e^{2\alpha\rho}\sqrt{-g}$, eq.\,\eqref{eq:RD} reproduces the usual reduction of the Einstein-Hilbert term, $\int \sqrt{-G}\,R_D=\int \sqrt{-g}\,[R-(\partial\rho)^2]$, the $\partial^2\rho$ term being a total derivative. For the higher-order terms in \eqref{eq:SD} this is no longer the case, since $\partial^2\rho$ then appears multiplied by $\rho$-dependent factors. Inserting \eqref{eq:RD} into \eqref{eq:SD}, we find a field-dependent scale
\begin{equation}
\label{eq:Mphi}
M(\rho) = M_{0}\, e^{\alpha \rho}=M_0\,e^{ \frac{\rho}{\sqrt{(d-1)(d-2)}}},
\end{equation}
and
\begin{equation}
\sqrt{-G}\,f(R_D) = \sqrt{-g}\,M(\rho)^2\,\hat f\!\left(\frac{R-2\alpha\partial^2\rho-(\partial\rho)^2}{M(\rho)^2}\right),
\qquad \hat f(x)=x+\sum_{n=2}c_n x^n\,.
\end{equation}
Expanding to quadratic order in $\partial\rho$ and integrating by parts, the resulting $d$-dimensional action becomes at tree level\footnote{Higher-loop effects can induce an additional curvature series whose characteristic scale is the Kaluza-Klein mass of the compact manifold \cite{Calderon-Infante:2025ldq}, and for irrelevant operators these can in some cases dominate over the tree-level reduction. In Type IIA/M-theory this occurs for example for terms of the form $D^n R^4$ for decompactification limits to $d=10$ \cite{Calderon-Infante:2025ldq}.}
\begin{equation}
\label{eq:SJ}
S_d = \frac{M_{\text{Pl}}^{d-2}}{2} \int d^d x \sqrt{-g} \left[\,M(\rho)^2\,\hat f(\hat R)\,-\,F\,(\partial \rho)^2\,+\,2\alpha\,\partial F\!\cdot\!\partial\rho\,\right],
\qquad \hat{R}\equiv\frac{R}{M(\rho)^2}\,,
\end{equation}
where $R$ denotes the $d$-dimensional Ricci scalar, and we have identified the lower-dimensional Planck mass as $M_{\text{Pl}}^{d-2} = 2\pi r\, M_{\text{Pl},D}^{d-1}$. Since $M(\rho)$ is now the scale controlling the curvature series, we can identify it parametrically with the species scale $\Lambda_{\rm sp}$ \cite{vandeHeisteeg:2023dlw}. We can do this because they have the same parametric dependence on the moduli of the theory, in this case the radion. However, we still assume, and to some extent require, that they can be numerically separated.

It has been conjectured \cite{Lee:2019wij, vandeHeisteeg:2023ubh, Castellano:2021mmx} that in string-theoretic setups the exponential coefficient $\gamma_{\rm sp}$ appearing in the species scale $\Lambda_{\text{sp}}\sim e^{\pm\gamma_{\rm sp}\rho}$ is necessarily bounded as
\begin{equation}
\label{eq:gspbound}
\frac{1}{\sqrt{(d-1)(d-2)}} < |\gamma_{\text{sp}}| < \frac{1}{\sqrt{d-2}}.
\end{equation}
The lower bound corresponds to a decompactification from $d$ to $d+1$ dimensions, as we have found in \eqref{eq:Mphi}, while the upper bound signals the emergence of a weakly coupled string. We then find that the exponential scaling of the potential and that of the species scale coincide only at the unique value $\gamma=1/\sqrt{(d-1)(d-2)}$. In short, the exponential decay rate corresponding to a potential arising from an infinite-order curvature expansion, eq.\,\eqref{eq:gammares}, can be matched to that of the species scale in a decompactification limit from $d$ to $d+1$ dimensions.

\subsection{Two-field Einstein-frame action}
\label{sec:iiib}

To pass to the Einstein frame we take $F=\hat f'(\hat R)$ as the independent variable, as in Section \ref{sec:ii}, and we perform the conformal transformation
\begin{equation}
g_{\mu\nu} \to \tilde{g}_{\mu\nu} = F^{2/(d-2)} g_{\mu\nu}\,.
\end{equation}
Here we have defined
\begin{equation}
\label{eq:sigmadef}
F = e^{c_d\,\sigma}\,,\qquad c_d\equiv\sqrt{\frac{d-2}{d-1}},
\end{equation}
so that $\sigma$ is the scalaron. For simplicity we take $M_{\text{Pl}}=1$. The Einstein-frame action takes the form\footnote{Here Einstein frame means that there are no terms depending on $R$ other than the Einstein-Hilbert term.}
\begin{equation}
\label{eq:S2f}
S = \int d^d x \sqrt{-\tilde g} \left[ \frac{1}{2} \tilde R - \frac{1}{2} \left((\partial \rho)^2+(\partial\sigma)^2-\frac{2}{d-1}\,\partial\rho\!\cdot\!\partial\sigma\right) - V(\rho,\sigma) \right],
\end{equation}
where the potential is
\begin{equation}
\label{eq:Vtwofield}
V(\rho,\sigma)=\frac{M_0^2}{2}\,e^{2\alpha\rho-d\alpha\sigma}\Big(\hat{R}\,F-\hat f(\hat{R})\Big)\,,\qquad F=\hat f'(\hat{R})=e^{c_d\sigma}.
\end{equation}
The metric on field space is
\begin{equation}
\label{eq:Kmat}
K=\begin{pmatrix} 1 & -\frac{1}{d-1}\\[3pt] -\frac{1}{d-1} & 1\end{pmatrix}
\end{equation}
so we can pass to a diagonal basis. Its eigenvectors are $\rho\pm\sigma$, with eigenvalues $\frac{d-2}{d-1}$ and $\frac{d}{d-1}$, so that the canonically normalized fields are
\begin{equation}
\label{eq:canfields}
\phi=\sqrt{\frac{d-2}{2(d-1)}}\,(\rho+\sigma)\,,\qquad
\psi=\sqrt{\frac{d}{2(d-1)}}\,(\rho-\sigma)\,.
\end{equation}
Rewriting the potential in these variables, the radion enters only through $\psi$ and in the plateau regime
\begin{equation}
\label{eq:Vcan}
V=\frac{M_0^2}{2}\,e^{\,\kappa \psi}\,\tilde{V}(\sigma)\,,\qquad
\kappa=2\alpha\sqrt{\frac{2(d-1)}{d}}=\sqrt{\frac{8}{d(d-2)}}\,,
\end{equation}
where $\tilde{V}(\sigma)=\hat{R}-\hat f(\hat{R})/F$ is the rescaled potential, which tends to unity at large $\sigma$, so that the potential is exponential in $\psi$ and flat in $\phi$. In the plateau regime the field $\phi$ is massless, so that $\ddot{\phi}+(d-1)H\dot{\phi}\simeq 0$ and its kinetic energy redshifts away as $a^{-2(d-1)}$.

\subsection{Lifting the potential}
\label{sec:iiid}

The potential \eqref{eq:Vcan} is a single exponential in $\psi$ and therefore has no stationary point, so that extremizing $\psi$ requires a second contribution with a different exponent. The exponent of a given term is fixed by its measure, according to whether it lives in the bulk and is integrated over the circle or is localized on a slice at fixed $x^d$. On the ansatz \eqref{eq:ansatz},
\begin{equation}
\label{eq:volfactors}
\sqrt{-G}=e^{(d\alpha+\beta)\rho}\sqrt{-g}=e^{2\alpha\rho}\sqrt{-g}\,,
\qquad
\sqrt{-g^{(J)}_{\text{ind}}}=e^{d\alpha\rho}\sqrt{-g}\,,
\end{equation}
where $g^{(J)}_{\text{ind}}$ is the metric induced on a slice of fixed $x^d$. The two differ by a factor of $e^{\beta\rho}$, which is the circumference of the circle. Passing to Einstein frame with $g=F^{-2/(d-2)}\tilde g=e^{-2\alpha\sigma}\tilde g$ multiplies both by $e^{-d\alpha\sigma}$, so that
\begin{equation}
\label{eq:weights}
\text{bulk:}\quad e^{2\alpha\rho-d\alpha\sigma}=e^{\,\kappa \psi}\,e^{-(d-2)\alpha\sigma}\,,
\qquad\qquad
\text{slice:}\quad e^{d\alpha(\rho-\sigma)}=e^{\frac{d}{2}\kappa \psi}\,.
\end{equation}
More generally, a term whose density in the Jordan frame scales as $e^{w\alpha\rho}$ contributes
\begin{equation}
\label{eq:weightrule}
e^{\alpha(w\rho-d\sigma)}=e^{\frac{w}{2}\kappa \psi}\,e^{\alpha(w-d)\sigma}\,,
\end{equation}
so that $w$ fixes both the weight in $\psi$ and a residual dependence on the scalaron. The bulk and slice sources correspond to $w=2$ and $w=d$, respectively. A source defined on the slice has an energy density $\mathcal{E}$ in units of its local frame, which cannot depend on $\psi$, and its contribution to the $d$-dimensional potential is therefore $e^{\frac d2 \kappa \psi}\mathcal{E}$, as would be the case for an extended object localized on the circle. Together with the bulk contribution, the potential takes the form
\begin{equation}
\label{eq:sV}
V=A_1(\sigma)\,e^{\,\kappa \psi}+A_2\,e^{\frac d2 \kappa \psi}\,,\qquad A_1=\frac{M_0^2}{2}\tilde{V}(\sigma)>0\,.
\end{equation}

Extremizing at $V>0$ requires $A_2<0$, so that the localized sector has to contribute negative vacuum energy. The stationary point is then a maximum, located at
\begin{equation}
e^{(\frac d2-1)\kappa \psi_*}=\frac{A_1}{\frac d2|A_2|}\,,\qquad
V_*=\frac{\frac d2-1}{\frac d2}\,A_1e^{\kappa \psi_*}\,\propto\,\tilde{V}(\sigma)^{\frac{d}{d-2}}\,,\qquad
\partial_\psi^2V\big|_{\psi_*}=-\frac{4}{d-2}\,V_*\,.
\end{equation}
The energy along the extremum is proportional to a power of $\tilde{V}$ and approaches a constant at the same exponential rate $\lambda$ as $\tilde{V}$ itself, so that the inflaton direction is as flat there as along $\psi=0$. Since $\psi$ is canonically normalized, its mass is given by
\begin{equation}
\label{eq:msridge}
m_\psi^2=\partial_\psi^2V\big|_{\psi_*}=-2(d-1)H^2\,,
\end{equation}
which is tachyonic and of order the Hubble scale, so that a de Sitter fluctuation $\delta \psi\sim H$ displaces the field within $\mathcal{O}(1)$ $e$-folds. This would then give a short period of accelerated expansion rather than a plateau of cosmological duration.

A minimum would require a third contribution, and any term growing faster in $\psi$ than the localized source would suffice. A Casimir energy is a natural candidate, since its scale is set by the size of the circle, which by \eqref{eq:weightrule} gives $w=d(d-1)>d$. It also acquires a dependence on $\sigma$, so that a full multifield analysis is beyond the scope of this work.

\subsection{Plateau potentials}
\label{sec:iiie}
We now assume that $\psi$ can in fact be fixed by the additional terms proposed in the previous section. Since it enters the potential only exponentially, a displacement simply implies a rescaling of the potential, and we can set $\psi\simeq 0$. Using \eqref{eq:sigmadef} and \eqref{eq:Mphi}, this is then equivalent to
\begin{equation}
\label{eq:matching}
F^{1/(d-2)}\,M_{\text{Pl}} \,=\, M(\rho)\,\simeq\,\Lambda_{\text{sp}}\,.
\end{equation}
We therefore assume that there exists a sector whose potential depends on the moduli only through $\psi$ and which fixes $\psi\simeq0$ with a mass $m_\psi\gtrsim H$. Everything in the remainder of this section, and the phenomenology of Section \ref{sec:iv}, is conditional on this assumption.

Because $\psi=0$ is a straight line in the flat field space \eqref{eq:Kmat}, the trajectory does not turn. No isocurvature perturbations are sourced, and integrating out the heavy direction affects the single-field description only at relative order $H^2/m_\psi^2$ \cite{Gordon:2000hv,Achucarro:2012sm}. The model then reduces to a single field with the potential \eqref{eq:Vcan} evaluated at $\psi=0$,
\begin{equation}
\label{eq:Vplateau}
S = \int d^d x \sqrt{-g}\left[\frac{1}{2}R-\frac{1}{2}(\partial \phi)^2-\frac{M_0^2}{2}\,\tilde{V}(\phi)\right],
\end{equation}
with $\tilde{V}$ determined by the curvature series and $\phi=\sqrt{2(d-2)/(d-1)}\,\rho$ the canonically normalized field along the locus, which follows from restricting \eqref{eq:Kmat} to $\sigma=\rho$. Therefore, the exponential rates quoted in Section \ref{sec:ii} differ by a factor of $\sqrt{(d-1)/2(d-2)}$, which in $d=4$ is $\sqrt{3}/2$. The bound \eqref{eq:lambdares} takes the form
\begin{equation}
    \lambda\leq1/\sqrt{2},
\end{equation}
independent of $d$. The exponential suppression of the potential has been cancelled against the exponential growth of the cutoff, and what survives is the plateau profile. In this regime the inflationary scale $H \sim \sqrt{V}$ remains parametrically below the species scale $M(\rho)$. Since we are looking at cosmological solutions one has an Einstein-frame curvature $R\simeq V$, which leads to the relation
\begin{equation}
\label{eq:Lspofr}
    \Lambda_{\text{sp}}\equiv M\simeq M_0 \left(\dfrac{M_0^2}{M_0^2-R}\right)^{\frac{1}{d-2}},
\end{equation}
where we have assumed $c_n\simeq \mathcal{O}(1)$. As such, there are two distinct scales, $\Lambda_{\text{sp}}$ and $M_0$. The former is the scale of quantum gravity, while the latter determines the cosmological constant, and they are related exponentially through the displacement of the inflaton. In what follows we analyze specific choices of $f(R)$ that yield viable plateau potentials, and we comment on their consistency with Swampland constraints.

\subsubsection{\texorpdfstring{$f(R) = R + M^2 \sum_n \frac{R^n}{M^{2n}}$}{f(R) = R + R\^{}n}}
For a series of curvature corrections in which every term is at the same order, $c_n\simeq 1$, we have
\begin{equation}
\label{eq:fRsum1}
    f(R)=R+M^2\sum_{n=2} \dfrac{R^n}{M^{2n}}=\dfrac{R}{1-R/M^2}\,,
\end{equation}
with $R<M^2$. Before imposing $\psi=0$, the scalar potential is
\begin{equation}
\label{eq:potfinal}
                V= \dfrac{1}{2} M^2 e^{-\frac{2}{\sqrt{(d-2) (d-1)}} \rho } \left(1-e^{-\frac{1}{2}\sqrt{\frac{d-2}{d-1}}\sigma}\right)^2,
\end{equation}
where the exponential rates satisfy eq.\,\eqref{eq:gammares} and eq.\,\eqref{eq:lambdares} with $q=0$. Imposing $\psi=0$, that is $M\to M_0\, e^{\alpha\rho}$ with $\sigma=\rho$, the exponential prefactor cancels and, after canonically normalizing $\phi$,
\begin{equation}
\label{eq:potRsum}
                \tilde {V}(\phi)= \dfrac{M_0^2}{2} \left(1-e^{-\frac{1}{2\sqrt{2}}\phi}\right)^2\,,
\end{equation}
where we note that the explicit dependence on the number of dimensions has cancelled between the rate of $\tilde{V}$ and the normalization of $\phi$. As such, one obtains Starobinsky-like models of inflation as the effective scalaron potential coming from an infinite series of curvature corrections, lifted by the exponentially varying species scale arising from the compactification of an internal circle.

\subsubsection{\texorpdfstring{$f(R)=R+M^2\sum_{n} n^q \frac{R^n}{M^{2n}}$}{f(R) = R + n\^{}q R\^{}n}}
As discussed in Section \ref{sec:ii}, the scalar potentials resulting from series with rational coefficients have very constrained asymptotic behavior. Whenever the curvature series has mostly positive terms and the coefficients are bounded as $c_n>1/n^2$, we will have a potential with overall exponential suppression that can be cancelled out after imposing $\psi=0$,
\begin{equation}
    \tilde{V}\simeq \dfrac{M_0^2}{2}\left(1-a\, e^{-\lambda \phi}\right)\,,
\end{equation}
with $\lambda$ obtained from eq.\,\eqref{eq:lambdares} together with the normalization of $\phi$. The same cancellation of the dimension dependence observed in \eqref{eq:potRsum} occurs for the whole family, and one finds
\begin{equation}
\label{eq:lamray}
\lambda =
    \begin{cases}
        \dfrac{1}{\sqrt{2}\,(2 + q)} & \text{for } q \geq -1\,, \\[6pt]
        \dfrac{1}{\sqrt{2}} & \text{for } -2 < q \leq -1\,,
    \end{cases}
\end{equation}
independently of the number of dimensions, and bounded above by $1/\sqrt2$. Given slow roll, inflationary phenomenology will only depend directly on the precise value of $\lambda$, so we briefly focus on series that have phenomenology close to that of the Starobinsky model. This corresponds to series with coefficients $c_n\simeq n^{q}$ with $-2\leq q\leq-1$, which saturate the bound in \eqref{eq:lamray} and are upper-bounded by
\begin{equation}
\label{eq:starolikepot}
    \tilde{V}(\phi)=\dfrac{M_0^2}{2}\left(1-e^{-\frac{1}{\sqrt{2}}\phi}\left(1+\tfrac{1}{\sqrt2}\phi\right)\right)\, \quad(q=-1)\,,
\end{equation}
\begin{equation*}
    \tilde{V}(\phi)\simeq\dfrac{M_0^2}{2}\left(1-e^{-\frac{1}{\sqrt{2}}\phi}\right)\, \quad(q=-2)\,,
\end{equation*}
where for the case of $q=-2$ we are assuming large $\phi$ to neglect subleading terms.
\begin{figure}
    \centering
\includegraphics[width=0.6\textwidth]{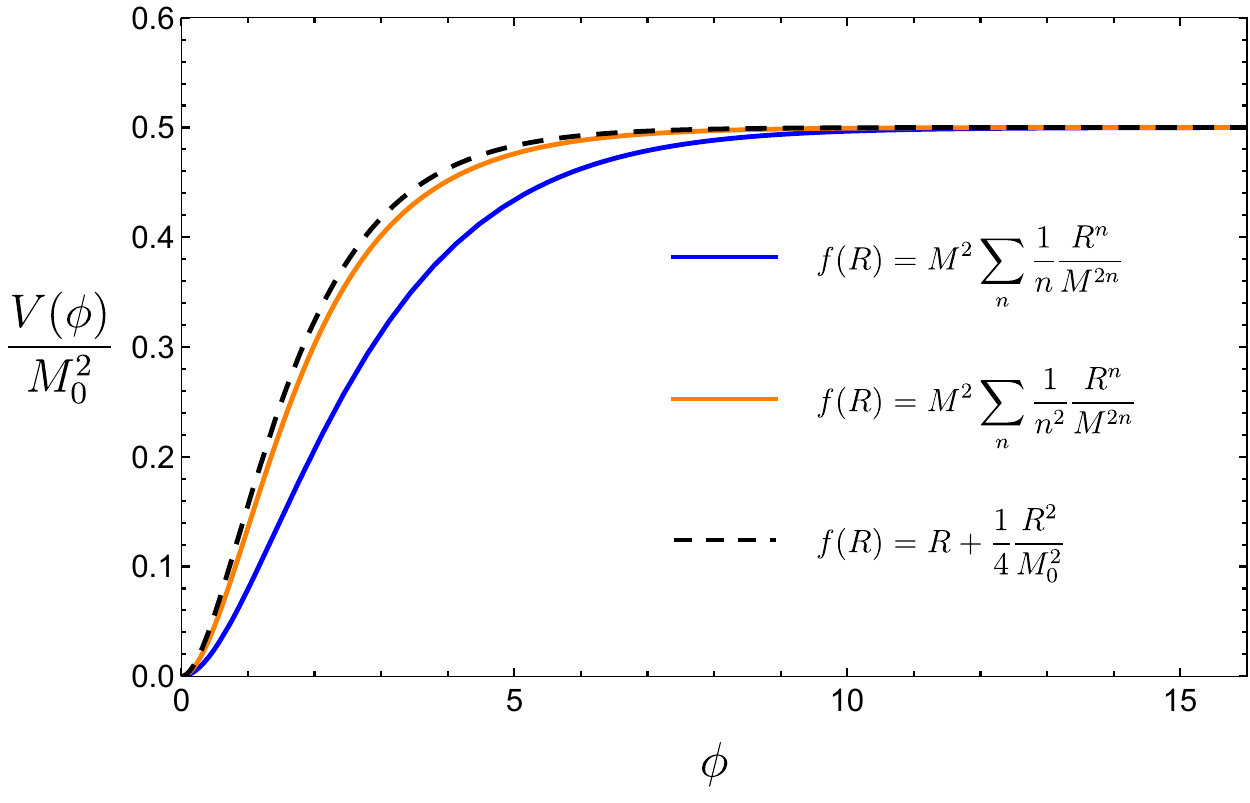}
 \caption{Potentials obtained from series of higher (scalar) curvature corrections to EH gravity after imposing $\psi=0$, for series with decreasing polynomial coefficients $c_n=n^q$. Blue: $q=-1$. Orange: $q=-2$, computed numerically. Black: Starobinsky model presented for comparison.}
    \label{fig:starocorrsums}
\end{figure}
These two potentials (see Fig.\,\ref{fig:starocorrsums}), and any resulting potential with coefficients $1/n^{2}<c_n<1/n$, will have equivalent phenomenology. This is discussed in Section \ref{sec:iv}.

One can ask which series would uplift to a Starobinsky-like potential of the form
\begin{equation}
    V(\rho)=\dfrac{1}{2}M^2 e^{-\sqrt{\frac{1}{2}}\rho}\left(1-e^{-\sqrt{\frac{1}{2}}\rho}\right)^2
\end{equation}
before imposing $\psi=0$. The condition can be expressed as a differential equation for $f(R)=\sum c_n (R/M^2)^n$,
\begin{equation}
    f'^2 \left(R-M^2\right) -f'\left(f-2M^2\right)=M^2\,,
\end{equation}
which then gives a recursive relation for the coefficients, after fixing $c_0=0, c_1=1$,
\begin{equation}
2 (n+1) c_{n+1} +\sum_{k=0}^{n} (n - k+1) c_{n - k+1} \left[ (k-1) c_{k} - (k + 1) c_{k+1} \right]  = 0\,.
\end{equation}
This leads to the coefficients expressed in closed form
\begin{equation}
    c_n=\dfrac{1}{(2n-1)\, 2^{2 n-1}}\binom{2 n}{n},\, \{c_n\}=(1,\frac{1}{4},\frac{1}{8}...)\,,
\end{equation}
which indeed satisfy $1/n^2 \leq c_n\leq 1/n$ and therefore belong to the class \eqref{eq:starolikepot}. We stress, however, that after canonical normalization this series does \emph{not} reproduce Starobinsky exactly. By \eqref{eq:lamray} its rate is $\lambda=1/\sqrt2$, whereas the Starobinsky model corresponds to $\lambda=\sqrt{2/3}$. The construction therefore yields a Starobinsky-\emph{like} plateau, distinguishable in principle from the original model.

It is worth mentioning that, provided $\psi$ can be stabilized, the setup proposed here evades the argument of \cite{Lust:2023zql}, where it was claimed that the scale controlling the $R^2$ term in Starobinsky inflation must necessarily be $M\simeq \Lambda_{\text{sp}}$. In the effective parameterization above one can have $M_0 \ll \Lambda_{\text{sp}}$, the two being related by \eqref{eq:Lspofr}.

Finally, we note that our analysis assumes an exponential dependence of the species scale on the canonically normalized field, as suggested by the SDC \cite{Ooguri:2006in}. This behavior signals that one is approaching the boundary of moduli (field) space, where an infinite tower of states becomes exponentially light. In our parameterization, the bulk of field space lies towards $\rho \gg 1$, where $\Lambda_{\mathrm{sp}}\,\simeq\,M_{\text{Pl}}$, and the exponential scaling is expected to break down \cite{vandeHeisteeg:2022btw}. On the other hand, at small $\rho\simeq 0$, where $\Lambda_{\mathrm{sp}}\!\simeq\! M_0$, the potential settles into its minimum and exits the plateau regime, signaling the end of slow roll.
The effective potential is thus expected to be reliable only within an intermediate region of field space, corresponding to a part of the plateau regime, where
\begin{equation}
\label{eq:window}
M_0\simeq H\ll \Lambda_{\text{sp}}\ll M_{\text{Pl}}\,.
\end{equation}

\section{Phenomenology}
\label{sec:iv}
We have seen that $f(R)$ theories of gravity in $d$ dimensions lead to potentials that decay exponentially fast as $V\propto e^{-2\gamma\sigma}$ when the curvature series is infinite, and that the cutoff controlling the series grows as $M=M_0e^{\gamma_{\rm sp}\sigma}$ after compactification on a circle, with $\gamma=\gamma_{\rm sp}=1/\sqrt{(d-1)(d-2)}$. Assuming $\psi$ can be taken constant as in Section \ref{sec:iiie}, the two effects cancel and the potential develops a plateau,
\begin{equation}
\label{eq:Vphigen}
    V(\phi)=V_0 \left(1- a \exp (-\lambda  \phi )\right)^n \,,
\end{equation}
with $\phi$ the canonically normalized inflaton. A potential of this form leads to the expected leading-order values of the scalar spectral tilt and tensor-to-scalar ratio
\begin{equation}
    n_s-1\simeq-\dfrac{2}{N_e} \,\text{ , }\, r\simeq\dfrac{8}{\lambda^2 N_e^2}\,,
\end{equation}
such that, at first order, the observables only depend on $\lambda$ and on the number of $e$-folds $N_e$. We would like to constrain the possible potentials by whether they can satisfy experimental bounds \cite{Planck:2018vyg,  BICEP:2021xfz}
\begin{equation}
    n_s-1= -0.035\pm 0.004 \,\text{ , }\, r<0.03\,.
\end{equation}
For potentials asymptotically reproducing the Starobinsky plateau, such as eq.\,\eqref{eq:starolikepot}, corresponding to curvature series with coefficients $1/n^2\leq c_n\leq 1/n$, eq.\,\eqref{eq:lamray} gives
\begin{equation}
    \lambda=\dfrac{1}{\sqrt{2}}\,,\qquad
    n_s-1=-0.033\,, \quad r= 0.0044\,,
\end{equation}
independent of $n$ and of the number of dimensions. These values should be compared with the Starobinsky model itself, for which $\lambda=\sqrt{2/3}$ and $r=0.0033$. The two are close but not identical, the difference being the normalization factor of Section \ref{sec:iiie}. Both easily satisfy current bounds, and the $30\%$ enhancement of $r$ relative to Starobinsky may be within reach of the next generation of B-mode experiments \cite{Abazajian:2019eic, LiteBIRD:2024wix}.

For a spectrum of coefficients obeying $c_n=n^q$ with $q\geq -1$ we found
\begin{equation}
    \lambda=\dfrac{1}{\sqrt{2}\,(2+q)}\,,\qquad
    n_s-1=-0.033\,, \quad r= (2+q)^2\, 0.0044\,,
\end{equation}
so that these series can fit the experimental bounds on $n_s$ and $r$ for $q\lesssim 0.6$. In particular the geometric series \eqref{eq:potRsum}, $q=0$, predicts $r=0.018$, comfortably within the current bound but well above the Starobinsky value.

In summary, the resulting potentials described in Section \ref{sec:iii}, arising from infinite curvature series with coefficients $c_n$ together with an appropriately varying UV scale, can fit inflationary bounds as long as the coefficients satisfy $1/n^2\lesssim c_n\lesssim n^{0.6}$.

\subsection{Exponentially decaying potentials}
\label{sec:iva}
If there is a slight deviation from $\psi=0$, parameterized as $\kappa \psi\simeq\beta \phi$, or if the exponential rates fail to match exactly because the curvature series is finite, the potential takes the form
\begin{equation}
\label{eq:Vphigen2}
    V(\phi)=V_0\, e^{\beta \phi}\, \left(1- \exp (-\lambda  \phi )\right)^n\,,
\end{equation}
with $\beta=2(\gamma-\gamma_{\rm sp})$ in the latter case. The parameter $\beta$ interpolates between the two regimes discussed above. Here $\beta=0$ is the plateau of Section \ref{sec:iiie}, while $\beta=\kappa$, with $\kappa$ given in \eqref{eq:Vcan}, is the unconstrained potential of Section \ref{sec:iiib}. Recalling the bounds on the exponential decay of the species scale,
\begin{equation}
   \dfrac{1}{\sqrt{(d-2)(d-1)}}  \leq|\gamma_{\rm sp}|\leq \dfrac{1}{\sqrt{d-2}}\,,\qquad
   \dfrac{1}{\sqrt{6}}  \leq|\gamma_{\rm sp\,,\,d=4}|\leq \dfrac{1}{\sqrt{2}}\,,
\end{equation}
while for a general series one has an exponentially decaying potential with
\begin{equation}
 0 \leq\gamma\leq\dfrac{1}{\sqrt{(d-2)(d-1)}}\,,\qquad
 0 \leq\gamma_{d=4}\leq\dfrac{1}{\sqrt{6}}\,,
\end{equation}
we find that these two exponents can coincide only for a single value, corresponding to a species scale arising from compactification on a circle and an infinite curvature series.

The spectral tilt and tensor-to-scalar ratio then receive corrections of the form \cite{Lust:2023zql}
\begin{equation}
    n_s-1 \simeq-\dfrac{2}{N_e}+\beta\lambda  \,\text{ , }\,  r\simeq\dfrac{8}{N_e^2\lambda^2}+\dfrac{8\beta}{N_e\lambda}\,,
\end{equation}
such that at leading order they have no dependence on $n$. For a number of $e$-folds $N_e=60$, and with $\lambda\leq1/\sqrt2$, experimental bounds on $n_s$ restrict $\beta$ as
\begin{equation} -0.008\lesssim\beta\lesssim   0.003\,.
\end{equation}
Since $\kappa=1$ in $d=4$, this means that $\psi$ must be fixed to better than about one percent. Equivalently, one can only consider a coupling $\gamma$ very close to $\gamma_{\rm sp}$. For a finite series of curvature corrections, eq.\,\eqref{eq:gammaexp} gives $\beta=-2/[\sqrt6\,(n-1)]$ in $d=4$, so that satisfying experimental bounds requires a leading curvature term $R^n$ with $n\geq \mathcal{O}(100)$.

\subsection{Curvature corrections in M-theory and \texorpdfstring{$f(R)$}{f(R)}}
\label{sec:ivb}

It is known that higher curvature corrections survive in M-theory and in 10-dimensional supersymmetric string theory. The first such correction in M-theory is given by
\begin{equation}
        S=\frac{M_{11}^9}{2} \int d^{11}x \sqrt{-g} \left(R + \frac{\left(4 \pi\right)^{2/3}}{(2 \pi)^4 \cdot 3^3 \cdot 2^{13}} \dfrac{1}{M_{11}^6}\left( \hat{t}_{8} \hat{t}_{8} \mathcal{R}^4 - \frac{1}{24} \epsilon_{11} \epsilon_{11} \mathcal{R}^4 \right)... \right) .
\end{equation}

Higher-order terms of the form $D^n R^p$ for $p \leq 6$, $n \leq 6$ have also been computed \cite{Green:1997di, Green:2005ba, Castellano:2021mmx, Calderon-Infante:2025ldq, Green:2010wi, DHoker:2014oxd}. However, even higher-order terms are no longer BPS and are not necessarily protected against radiative corrections. We will assume a generic series of curvature corrections and attempt to infer a relation for their coefficients, independently of the reduction of the setup considered in Section \ref{sec:iii}.

Minkowski space $\mathbb{R}^{1,10}$ is a known solution of M-theory, but $AdS_p \times S^{11-p}$ solutions preserving maximal supersymmetry have also been studied \cite{Aharony:1998rm}. In particular, these solutions are locally symmetric spacetimes with constant curvature, and the Riemann tensor can be written in block-diagonal symmetric form, despite the spacetime not having the maximal number of Killing vectors. Since these spaces typically do not exhibit scale separation, one has $R_{\text{AdS}}=R_{\text{KK}}=R$, such that every curvature invariant can be expressed in terms of $R$. Motivated by this, we assume that M-theory allows for spaces that satisfy:
\begin{equation}
    R_{\mu\nu\rho\sigma}=\dfrac{R}{110}\left(g_{\mu\rho}g_{\nu\sigma}-g_{\mu\sigma}g_{\nu\rho}\right)\,.
\end{equation}

Under this ansatz, the $\mathcal{R}^4$ term simplifies to
\begin{equation}
    \frac{\left(4 \pi\right)^{2/3}}{(2 \pi)^4 \cdot 3^3 \cdot 2^{13}} \dfrac{1}{M_{11}^6}\left( \hat{t}_{8} \hat{t}_{8} \mathcal{R}^4 - \frac{1}{24} \epsilon_{11} \epsilon_{11} \mathcal{R}^4 \right)\simeq \dfrac{R^4}{M_{11}^6}.
\end{equation}

Given the existence of an infinite tower of curvature corrections, as argued in \cite{vandeHeisteeg:2022btw, vandeHeisteeg:2023ubh}, the effective action takes the form:
\begin{equation}
        S=\frac{M_{11}^9}{2} \int d^{11}x \sqrt{-g} \left(R +\sum_n^N c_n \dfrac{R^n}{M_{11}^{2n-2}}\right).
\end{equation}
We are working in the low-energy limit of M-theory, namely 11-dimensional supergravity, so we should generally expect such an infinite series to be an \textit{asymptotic series}. Here, we assume that the leading term in the series is of order $N > 11/2$, so that the resulting potential decays exponentially. Depending on the nature of the coefficients, the low-energy sector of the theory can then vary considerably. For $c_n > 1/n^2$, we obtain after going to the Einstein frame:
\begin{equation}
            S=\frac{M_{11}^9}{2} \int d^{11}x \sqrt{-g} \left(R -(\partial\sigma)^2-V_0 e^{-\gamma\sigma}\right),
\end{equation}
with $V_0\simeq M_{11}^{11}$ and $\gamma > 0$.

Several issues arise in this case. In particular, one can formally take a Minkowski limit by expanding $\sigma(x)=\sigma_0+\delta\sigma(x)$ and taking $\sigma_0$ to infinity, where the exponential potential vanishes. However, once the potential drops below any other scale in the theory, $\delta\sigma$ effectively behaves as a modulus. This is problematic from the standpoint of M-theory, where the moduli space is expected to be trivial \cite{Cremmer:1978km, Horava:1996ma}. Indeed, we have started from M-theory and ended up with the same theory plus an additional massless field. Furthermore, this being an infinite-distance limit suggests that we should expect a further decompactification or some other exotic limit, which would perhaps indicate that the curvature series cannot be resummed in the way we have described.

In fact, none of these issues arise when the coefficients obey $c_n < 1/n^2$, since $\sigma$ is not allowed to run to infinity\footnote{$-\infty$ is technically allowed, but an exponential wall prevents $\sigma$ from approaching this limit.}. For illustrative purposes, we assume coefficients of the form $c_n = n^q$. For $q<-2$ there is a single Minkowski minimum at $\sigma = 0$, and the field gains a mass:
\begin{equation}
    m_{\sigma}= \sqrt{\dfrac{9}{10
    \,2^{q+2}}}M_{11}\,.
\end{equation}

In this case, the scalar mass is of the order of the cutoff, so the field can be integrated out, and one recovers the expected content of M-theory in flat space. We then propose that at least one of the following must hold in M-theory:
\begin{itemize}
    \item One cannot continuously deform a maximally symmetric universe to flat space.
    \item There is no optimal truncation at order $\mathcal{R}^N$ with $N>11/2$. In this sense, there is no regime in which one can consider the gravitational series as finite at order $N>11/2$.\footnote{The first term to break all supersymmetry is $\mathcal{R}^8$, so it is perhaps expected that the series starts diverging once the terms are no longer BPS-protected.}
    \item The coefficients of the gravitational expansion must asymptotically satisfy $c_n < 1/n^2$.
\end{itemize}

 We are also not aware of any top-down derivation of the coefficient spectra considered here. If anything the expectation runs the other way, since the perturbative expansion of string theory is itself asymptotic, with the genus $g$ contributions growing as $(2g)!$ \cite{Gross:1988ib,Shenker:1990uf}, so that a series with polynomially decaying coefficients and a finite radius of convergence is not what a complete ultraviolet description would be expected to produce. In consistent effective theories coupled to gravity, such as ten-dimensional supergravities descending from string theory, the infinite tower of higher-derivative terms is expected to form an asymptotic series rather than a convergent one, and non-BPS-protected terms generally receive radiative corrections. Together with the argument above this would imply that a convergent curvature series evaluated on a maximally symmetric background is not compatible with known UV complete theories. The results of Section \ref{sec:iii} should therefore be read as a property of the class of $f(R)$ theories we study, and not as a statement about M-theory.

\section{Conclusion}
\label{sec:v}
In this work we have analyzed how higher curvature corrections in $f(R)$ gravity give rise to scalar potentials and how these connect to the notion of a dynamical species scale. Starting from general considerations of $f(R)$ models in arbitrary dimensions, we found that the generic outcome of infinite curvature series is an exponentially decaying potential at large field values. We then established a connection between these potentials and the exponential behavior of the species scale.
Reducing the $f(R)$ theory on a circle, the radion and the scalaron span a flat, non-diagonal field space and the radion enters the potential only through a single combination $\psi$, so that the potential is exponential in $\psi$ and flat along the orthogonal direction $\phi$. The resulting two field potential has no critical point in $\psi$, so that, without additional terms, $\psi=0$ is not a solution of the equations of motion and the system is driven along $\psi$. We argue that the potential can be extremized in $\psi$ by a combination of localized and bulk sources, with $m_\psi^2=-2(d-1)H^2$. Since $\psi$ is tachyonic it remains in place for at most an $\mathcal{O}(1)$ number of $e$-folds. Stabilizing $\psi$ requires a third contribution, which we find is possible whenever its weight in $\psi$ exceeds that of the localized source, although we do not identify the sector that provides it. Under the assumption that it can be stabilized, the exponential decay of the potential is compensated by the exponential growth of the species scale, and the inflationary scale remains parametrically below the cutoff throughout the relevant field range, ensuring the consistency of the effective description even in the presence of an infinite tower of corrections.

From a phenomenological perspective, the resulting potentials are reminiscent of Starobinsky inflation, with predictions for the tensor-to-scalar ratio at $r=16/N_e^2$, instead of $r=12/N_e^2$. Their predictions are essentially determined by the asymptotic structure of curvature series and the number of $e$-folds. The precise value of the exponential rate $\lambda$ is then highly constrained, and may come within experimental reach in the coming decade \cite{Abazajian:2019eic, LiteBIRD:2024wix}.

We also commented on possible implications in string and M-theory, where series of curvature corrections are expected to arise \cite{Green:2005ba, Green:1997di}. While the structure and convergence of such series remain uncertain, we have argued that the resulting potentials can be problematic if the coefficients do not decay sufficiently fast, for instance slower than $c_n \sim 1/n^2$.

To conclude, we have highlighted a simple mechanism linking higher curvature corrections, the species scale, and plateau-like inflationary potentials. It is worth mentioning that alongside curvature series weighted down by the species scale there is also a series of higher-loop corrections weighted down by the KK scale of the compact space, and these have been shown to be the dominant contribution in some cases \cite{Calderon-Infante:2025ldq}. This remains a subject for future work, as it would be interesting to see if the mechanism we have outlined could be obstructed by higher-loop corrections.

\section*{Acknowledgements}
We would like to thank A. Herraez and A. Paraskevopoulou for useful conversations and comments on this draft, as well as D. Lüst, A. Hell, I. Basile and S. Schreyer for helpful discussions. We would also like to thank M. Scalisi for helpful comments, as well as the Department of Physics and Astronomy ``Ettore Majorana'' at the University of Catania for their kind hospitality during the completion of part of this work.
\appendix
\section{Higher curvature terms in quasi-de Sitter}
\label{sec:appi}

The restriction to $f(R)$ is an assumption we make throughout, and in this appendix we show the weaker statement that on a quasi-de Sitter background the invariants we have dropped are suppressed by powers of $\varepsilon$ and $H/\Lambda$.

For an FLRW background with scale factor $a(t)$ and Hubble parameter $H=\dot a/a$, the Ricci scalar in $d$ dimensions is
\begin{equation}
    R \,=\, 2(d-1)\dot H + d(d-1)H^2=d(d-1)H^2\!\left(1 - \frac{2}{d}\,\varepsilon\right),
\end{equation}
with the slow-roll parameter defined as
\begin{equation}
    \varepsilon \equiv -\frac{\dot H}{H^2}.
\end{equation}
During a slow-roll phase, we have $\varepsilon\ll 1$, and the background is (nearly) maximally symmetric, with
\begin{equation}
    R_{\mu\nu} \simeq \frac{R}{d}\, g_{\mu\nu}+\mathcal{O}(\varepsilon),
    \qquad
    R_{\mu\nu\rho\sigma} \simeq \frac{R}{d(d-1)}\big(g_{\mu\rho}g_{\nu\sigma}-g_{\mu\sigma}g_{\nu\rho}\big)+\mathcal{O}(\varepsilon),
\end{equation}
so that any local invariant made out of $k$ Riemann tensors satisfies
\begin{equation}
    \underbrace{\mathcal{R}\cdots\mathcal{R}}_{k\ \text{times}}
    \,\propto\, R^{k} \,+ \mathcal{O}(\varepsilon)\,.
\end{equation}

We can also consider operators with higher derivative orders
\begin{equation}
    \mathcal{O}_{m,n} \sim (\nabla^{2})^{m}\, R^{\,n}.
\end{equation}
Using $\nabla_\alpha R = \mathcal{O}(\varepsilon H^3)$ and $\nabla^2 R=\mathcal{O}(\varepsilon H^4)$ on the background, each pair of derivatives brings an extra suppression by $\varepsilon$
\begin{equation}
    (\nabla^{2})^{m} R^{\,n}
    \,\sim\, \varepsilon^{m}\, H^{2(n+m)} 
    \,\sim\, \varepsilon^{m}\, R^{\,n+m}.
\end{equation}

Then, for a gravitational EFT of the form
\begin{equation}
    \mathcal{L}
    \,=\, \frac{1}{2}R \,+\, \sum_{n\ge 2}\frac{c_{0,n}}{\Lambda^{2n-2}} R^{n} 
    \,+\, \sum_{m\ge 1,\, n\ge 1}\frac{c_{m,n}}{\Lambda^{2(n+m)-2}} (\nabla^{2})^{m} R^{n}
    \,+\,\cdots,
\end{equation}
with $\Lambda$ the UV cutoff, we will have that on a slow-roll background,
\begin{equation}
    \frac{\mathcal{O}_{m,n}}{R^{n}}
    \,\sim\, \varepsilon^{m}\left(\frac{H}{\Lambda}\right)^{2m}.
\end{equation}
Thus, as long as
\begin{equation}
    \varepsilon \ll 1,
    \qquad
    \frac{H}{\Lambda} \ll 1,
\end{equation}
the leading derivative corrections will be parametrically suppressed relative to the $R^{n}$ terms. It is relevant to mention that this amounts to evaluating the action on a set of solutions, and that the resulting $f(R)$ theory is not an EFT in the traditional sense, but rather a phase-space reduction.

\section{Asymptotic Behavior of Infinite Series}
\label{sec:appii}
In this appendix we discuss the behavior of infinite series that define $f(R)$ near their finite radius of convergence.
\subsubsection*{Polynomial coefficients}
Consider coefficients of the form \(c_n = n^q\). The associated power series has radius of convergence $|R|<M_0^2$ for $q \ge -1$, and $|R|\le M_0^2$ for $q < -1$, since the series converges also at the boundary point. In these regimes one finds closed forms:
\begin{equation}
  f(R)=\frac{M_0^2\, P_{q}\!\left(\frac{R}{M_0^2}\right)}{\left(1 - \frac{R}{M_0^2}\right)^{q+1}},
  \qquad q \ge 0,
\end{equation}
\begin{equation}
\label{eq:secondseries}
  f(R)=M_0^2\, \mathrm{Li}_{-q}\!\left(\frac{R}{M_0^2}\right),
  \qquad q \le -1,
\end{equation}
where \(P_q\) is a polynomial of degree \(q\) and \(\mathrm{Li}_\nu\) denotes the polylogarithm of order \(\nu\).

The series in \eqref{eq:secondseries} in fact converges for $R<M_0^2$ and also at the boundary $R=M_0^2$. However, only for $q>-2$ does the limit $\sigma\to\infty$ correspond to $R\to M_0^2$. For $q\le -2$, the potential is supported only up to a finite field range in $\sigma>0$.

\subsubsection*{Exponential coefficients}

If $c_n = a\,e^{\mu n}$ with fixed $a,\mu \in \mathbb{R}$, then
\begin{equation}
      \sum_{n\ge1} c_n \Big(\frac{R}{M_0^2}\Big)^n
  \,=\,
  a \sum_{n\ge1} \Big(e^\mu \frac{R}{M_0^2}\Big)^n
  \,=\,
  a \sum_{n\ge1} \Big(\frac{R}{M_0'^2}\Big)^n,
  \qquad M_0'^2 \equiv M_0^2 e^{-\mu}.
\end{equation}

Thus an exponential weight can be absorbed into a rescaling of the reference scale $M_0$. The large-$R$ behavior is unchanged, and only the effective radius of convergence shifts.

\subsubsection*{Super-exponential coefficients and optimal truncation}

If $c_n$ grows faster than any $e^{\mu n}$ (e.g. factorial growth), the radius of convergence is zero and the series is asymptotic, with optimal truncation at order
\begin{equation}
      \Big|c_{N+1}\Big(\dfrac{R}{M_0^2}\Big)^{N+1}\Big|
  \,\approx\,
  \Big|c_{N}\Big(\dfrac{R}{M_0^2}\Big)^{N}\Big|.
\end{equation}

In such cases, one may consider truncating the series at some finite order $n \leq N$ \cite{Goroff:1985sz, Bender:1999}.

\bibliography{refs}
\bibliographystyle{JHEP}
\end{document}